# Appearance of lines due to K ≠ 0 oscillations in lattice vibrations spectra of paradibrombenzene nanoparticles

M.A.Korshunov

*L.V. Kirensky Institute of Physics Siberian Branch of the RAS, 660036 Krasnoyarsk, Russia*

E-mail: kors@iph.krasn.ru

We measured low-frequency Raman spectra of paradibromobenzene nanoparticles. With the reduction of the nanoparticles size frequencies of lines become smaller and the doubling of the most intensive lines takes place. Additional lines with increasing intensity appear in the spectrum. We did structure calculations for nanoparticles by the method of molecular dynamics and obtained histograms of lattice vibrations spectra by the Dyne's method. Calculations have shown that doubling of lines is due to the appearance of nanoparticles vibrations from borders of the Brillouin zone in the spectrum. Additional lines are caused by vibrations with $\mathbf{K} \neq 0$. They have mixed orientation-translational character (that is evident from the eigenvectors of vibrations) that affects intensity of these lines. Upon reduction of the particles sizes change of vibration eigenvectors of molecules occurs $\mathbf{K} \neq 0$. The greatest changes observed for a 20 cm$^{-1}$ paradibromobenzene line.

## Introduction

At calculation of a spectrum of frequencies lattice fluctuations of a molecular crystal usually the model of an infinite crystal is replaced with model with cyclic boundary conditions. The knowledge of own vectors allows to define intensity of lines of a spectrum lattice fluctuations [1]. Change of the form of fluctuations changes tensor polarizability that affects change intensities lines. For an infinite crystal in a spectrum fluctuations are shown at wave vector $\mathbf{K} = 0$ (and dispersive curves have a continuous appearance) thus for an ideal crystal there is no mixture libration and transmitting fluctuations. At libration fluctuations there is a change of polarizability of molecules. Intensity of lines is connected with change of polarizability of molecules. Therefore in a spectrum of combinational dispersion of light of small frequencies the lines caused ориентационными by fluctuations of molecules (basically round the main moments of inertia) have the big intensity, than transmitting fluctuations as for these fluctuations there is no change of polarizability of molecules.

But if the sizes of a crystal decrease, dispersive curves suppose decisions corresponding to discrete values $\mathbf{K}$. Besides in nanocrystals because of restriction of the sizes there is an uncertainty of value of a wave vector $\mathbf{K}$ and fluctuations from various points of a zone of Brilljuena [2] are shown. The effect phonon confinement (model of spatial restriction фононов) can be shown as in displacement in low-frequency area and dissymetric broadening a fundamental fashion depending on the size and a kind nanoparticlesы [3-5]. Displacement of frequencies of lines in low-frequency area it is marked in work [6] and in a spectrum organic molecular nanocrystals [7]. The knowledge of the dispersive law allows to investigate soft fashions and the mechanism of phase transition, to define macro- both microscopic characteristics of substances and thermodynamic functions. Dispersive curves also help with research and phonon crystals [5]. In a spectrum lattice fluctuations for infinite crystals we observe spectrum lines at **K =0,** dispersive curves can be calculated математически. But in nanocrystals the spectrum is observed at different values **K**. It allows to investigate behaviour of dispersive curves, as well as at research of dynamics of a molecular crystal, a method of dispersion of neutrons. Therefore it is represented interesting to investigate spectrum change lattice fluctuations of molecular crystals with change of their size.

## Experiment

For organic nanoparticles are often used crushing raw material [2.11]. The second method used is sputtering on glass material being studied. In both methods, the evaporation decreases when the particle size, which allows to obtain the desired particle size. In this case, both methods were used. Dimensions nanoparticles were determined by an electron microscope. Thereafter, the record Raman spectra nanoparticles spectrometer Jobin Yvon T64000.

Preparation of Raman spectra of low-frequency nanoparticles paradibrombenzola changing their size from ~ 500nm do100nm. In the spectra of nanoparticles is also observed the appearance of additional lines, the intensity of which increases with decreasing size of nanoparticles. A detailed study of the spectra found that the spectrum

To explore this issue in the present work was a study of the experimental Raman spectra lattice vibrations paradibrombenzola nanoparticles and comparison with calculations of the lattice dynamics of the crystal. Paradibrombenzola single crystals have been well studied by various methods. There interpretation of lattice vibrations [8-10].

of nanoparticles paradibrombenzola observed doubling of the most intense lines of the spectrum is shown in Table 1.Razdvoenie lines observed for particles smaller than 500 nm. Most simply observe the changes in the spectrum in separate spaced lines. This first line is the most intense line associated with orientation in the field of molecular vibrations 20 $cm^{-1}$ in paradibrombenzole. The spectrum of a single line of about 20 $cm^{-1}$ and around additional lines of this line is resized nanoparticles ( 500 nm (a), 200 (b), 130nm (c) and 100 nm (d)) is shown in Fig.1. Line rotational vibrations of air molecules were subtracted from the original spectrum. As seen from the table while reducing the particle size to about 100 nm, the frequency number of lines decreased. Thus there is an asymmetry of the first line of the spectrum increases with decreasing size nanoparticlesy.

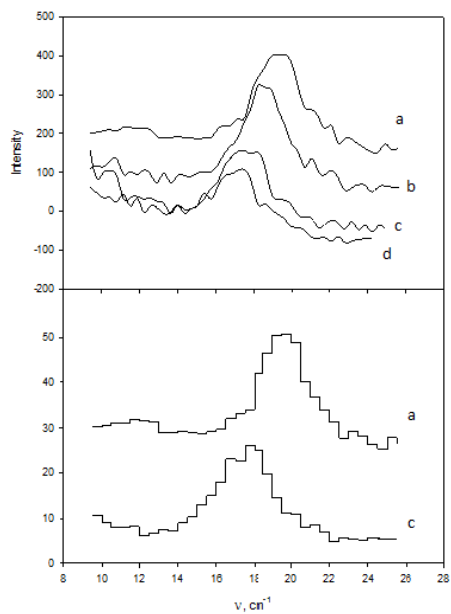

**Figure 1.** The line spectrum 20 cm-1 paradibrombenzola lattice vibrations at a particle size of from about 500nm to 100nm. Graph lower the smaller nanoparticles. Calculated histogram corresponding to the experimental spectrum (a, c).

**Table 1.** Doubling the line frequency (if the size of 200 nm nanoparticles) paradibrombenzola ν (cm$^{-1}$).

| парадибромбензол | | |
|---|---|---|
| 500нм | 200нм | |
| Частота(ν) | | |
| 18 | 16.8 | 17.5 |
| 35 | 33.0 | 34.7 |
| 36 | 35.5 | 36.3 |
| 39 | 37.2 | 38.0 |
| 90 | 87 | 89.0 |
| 97 | 93 | 95.0 |

### Details of calculations

Using the method of molecular dynamics and the absorption spectra lattice vibrations can be studied at the level of structure nanoparticlesy motion of individual molecules [12]. In the molecular structures, in contrast to structures consisting of atoms, the potential energy depends not only on the relative locations of the centers of gravity of the molecules, but also the orientation of each of the molecules, which significantly increases the calculation time. Molecular structure was assumed absolutely rigid. Interaction potential was chosen in the form ( 6 -EXP) [ 13]. In calculating the coefficients used in the interaction potential obtained earlier [ 14] in which the calculated spectrum of lattice vibrations of single crystals of paradichlorobenzene, paradibrombenzola and their solid solutions match the values of the frequency lines of experimental spectra obtained by lattice vibrations polarization studies investigated single crystals. To calculate the coordinates and velocities of the molecules by the method of molecular dynamics Verlet algorithm was used in the form of high-speed [15]. For nanoparticles paradibrombenzola found that with decreasing particle size increases the lattice parameters. Found using the structure were calculated spectra of lattice vibrations by the method of Dean [16]. On the basis of calculations yielded histograms that show the probability of spectral lines in the selected frequency range.

Calculated histogram spectra in cm$^{-1}$ 20 shown in Figure 1. Histograms were calculated for different values of the wave vector k. Coefficients found for this dynamic matrix multiplied by the phase factors for the spatial grid of values of k This found that the oscillations appear in the spectrum of the mixed type ( orientation-translational ). Table 2 shows an example of two oscillation frequencies when **K ≠ 0**. And their own vector (the first three are related to the orientation of the molecule vibrations and second with translational) as we can see, when **K ≠ 0** fluctuations are mixed orientation- translational.

**Table 3.** Frequency lines paradibrombenzola ν (cm$^{-1}$) and their eigenvectors with **K ≠ 0**.

| ν    | Θu    | Θv    | Θw    | x     | y     | z     |
|------|-------|-------|-------|-------|-------|-------|
| 18.7 | 0.3   | 0.59  | 0.156 | 0.035 | 0.245 | 0.305 |
| 10   | 0.054 | 0.347 | 0.277 | 0.809 | 0.367 | 0.103 |

Intensity, which is usually higher than the intensity of translational vibrations. This is caused by the appearance of additional lines in the spectrum. Also, we calculated the changes eigenvectors oscillations at different values of the wave vector, depending on the size of the crystals (Figure 2) and for different frequencies (Figure 3). Particle size in the calculations was taken equal to 150 nm, 300nm and endless crystal. Size nanoparticles is reflected in the magnitude of the wave vector. With increasing particle size is less than the wave vector changes.

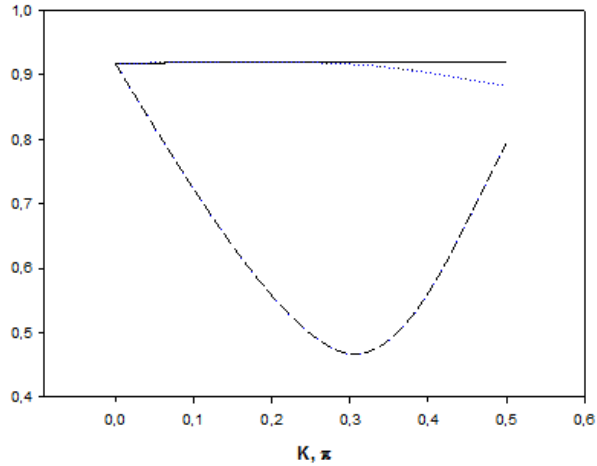

**Fig. 2.** Changing eigenvectors Θv line 20 for oscillation cm$^{-1}$ paradibrombenzole when resizing nanoparticles.

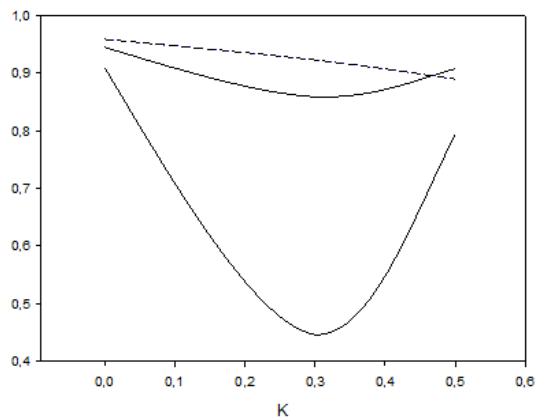

**Fig. 3.** Changing the waveform depending on the wave vector of some lines in paradibrombenzole for orientational fluctuations lower graph line for 20 cm$^{-1}$

## Discussion

The results of the study of Raman spectra of low frequencies paradibrombenzola nano particles. In the spectrum of a single crystal lattice vibrations paradichlorobenzene at **K = 0** should be observed six intense lines due to rotational oscillations of the molecules around the axes principal moments of inertia. First, the most intense line in the spectra of para-substituted benzene is associated with orientation fluctuations, and it is located separately from the rest of the spectral lines. So celebrate the change of its parameters with decreasing particle size more than just on this line. Figure 1 shows the changes in the first 20 lines cm$^{-1}$ paradibrombenzola. Paradibrombenzol was chosen because it has no phase transitions as seen with decreasing size of nanoparticles paradibrombenzola observed doubling of the line spectrum ( Figure 1). Additionally, the spectrum shows the appearance of additional lines of low intensity value of which increases with decreasing size of nanoparticles. The dispersion curve has a discrete character and we see the spectrum for different values of the wave vector of this causes the appearance of additional spectrum discretely spaced lines ( Figure 1). At the boundaries of the Brillouin zone at **K = 0** and **K = π** fluctuations are mainly to the orientation, this causes their greater intensity than the intensity of the lines at other K. These lines from the Brillouin zone boundaries cause a doubling of the spectrum. Magnitude of the splitting is equal to the frequency difference in the dispersion branches limits K. While at other waveform is mixed (Table 3). Therefore observed in the spectrum two intense line for each of the branches of oscillations with the limit values of K and a number of less intense lines at different values of **K**. When **K ≠ 0** the greatest change observed waveforms for line 20 cm$^{-1}$ ( lower graph in Figure 3) similar changes found in the calculations for the translational modes. When reducing the size of the crystal changes observed increase waveforms for different values of the wave vector ( Figure 2 shows the variation of the waveform for paradibrombenzola line at 20 cm$^{-1}$ at reducing the size of the nanocrystals ). Particle size in the calculations was taken equal to 150 nm, 300nm and endless crystal. With increasing particle size is less than the wave vector changes. It also depends on the frequency of oscillation.

As seen from Table 1 and Figure 1 with a decrease in particle size to about 100 nm, the frequency number of lines decreased. Thus there is an asymmetry of the first line of the spectrum increases with decreasing size nanoparticlesy. Change in the spectrum with decreasing size nanoparticles associated calculations showed a rearrangement of the lattice structure and change its parameters. The calculations of the frequency spectrum by the method of Dean showed that if the count range for different parts of the discrete dispersion curve is observed doubling of spectral lines. In this case oscillations are orientation- translational character.

## Conclusion

Carried out in this paper studies the changes in the spectra of nanoparticles paradibrombenzola decrease in their size showed that in all cases there is a doubling of the most intense lines of the spectrum. Particularly noticeable for the line spectrum in the 20 cm$^{-1}$. The experimental Raman spectra of low-frequency nanoparticles paradibrombenzola when resizing from 500 nm to 100 nm showed that a decrease in their size frequency values fall lines. When reducing the size of the nanoparticles paradibrombenzola ~ 500 nm to ~ 130 nm, there is an increase of the distance between the lines observed in the 20 cm$^{-1}$ to 0.5 to 1 cm$^{-1}$. The spectrum has an appearance of additional lines. To the resultant structure nanoparticles were calculated spectra histograms lattice vibrations method Dean. Calculations have shown that the spectrum there are additional lines due manifestation of oscillations with **K ≠ 0**. They have the form of orientation- translational vibrations that affect their intensity. Doubling of lines in the spectrum due to the manifestation nanoparticles oscillations with the Brillouin zone boundaries. With decreasing particle size increase observed changes in the shape of molecular vibrations with **K ≠ 0**. Greatest changes occur for line 20 cm$^{-1}$ paradibrombenzola. Thus, the nanoparticles are observed in the spectrum of the line spectrum at **K ≠ 0**. Particle size in the calculations was taken equal to 150 nm, 300nm and endless crystal. Size nanoparticles is reflected in the magnitude of the wave vector. With increasing particle size is less than the wave vector changes.